\documentclass[12pt]{article}
\usepackage{amsmath}
\usepackage{graphicx,psfrag,epsf}
\usepackage{enumerate}
\usepackage{natbib}
\usepackage{amsmath}
\usepackage{amssymb}
\usepackage{algorithm}
\usepackage{bbm}
\usepackage{fancyhdr}
\usepackage{amsfonts}
\usepackage{subfig,xcolor}
\usepackage{hyperref}
\usepackage{amsthm}
\usepackage{wrapfig}
\usepackage{mathtools}  
\usepackage{enumitem}
\mathtoolsset{showonlyrefs}
\usepackage[noend]{algpseudocode}

\usepackage{hyperref}
\usepackage{multirow}

\begin{document}

\def\spacingset#1{\renewcommand{\baselinestretch}%
{#1}\small\normalsize} \spacingset{1}

\title{\bf Bayesian Hierarchical Emulators for Multi-Level Models: BayHEm}
  \maketitle
  
\centering \author{Louise Kimpton, James Salter, Xiaoyu Xiong, Peter Challenor \\
University of Exeter, Department of Mathematics and Statistics, Exeter, UK}
  

\spacingset{1}

\section{Introduction}

Computer codes play a crucial role in describing complex physical phenomena across various scientific and engineering disciplines. With advancements in physics and computer science, simulators have become more intricate, making sensitivity analysis, uncertainty quantification, and optimization computationally challenging. To address this, surrogate models, like Gaussian process regression, are constructed to provide a fast approximation of the complex code, mitigating computational costs. 

These complex computer codes often exist in a hierarchical structure, running at various levels of complexity, ranging from the very basic to the sophisticated. Different levels of the model may be produced by varying resolutions, either in terms of time or spatial resolution. Alternatively, models can be executed at different complexity levels, allowing for the removal or imposition of certain components, assumptions, or features, thus creating a hierarchy of simpler models. The top levels of these computer codes come with a high computational cost, limiting the number of feasible runs. While our primary interest lies in understanding the most complex levels, cheaper runs at different levels of the hierarchy contain valuable information that aids in inferring underlying features of the entire system.

A widely adopted approach for handling complex computer codes within a multi-fidelity framework was introduced by \cite{Kennedy2000}.  The autoregressive model framework states that the computer code at level $l$ is equal to the computer code at level $l-1$ plus a corrector term which then evaluates the difference between these two neighbouring levels.  The strength of this approach lies in simplifying the learning process about the top-level code by focusing on two more manageable functions. This simplification allows for efficient learning, requiring fewer runs of the computationally intensive top-level code, and helps minimize uncertainty in the predictions. The method also contains a Markov-like property such that if the model at level $l-1$ is known at input parameter $x$, then no more can be learned about the model at level $l$ at $x$ from any other input parameter $x'$ at level $l-1$.

In this paper, we introduce a new approach to building Gaussian processes for complex multi-level computer codes.  In order to make sure that all information is included to model the top level, we define that the posterior distribution for level l-1 becomes the prior distribution for the model at level l. This means that we improve on the Markov-like property from Kennedy and O'Hagan, to now state that any inference made for level l can learn information from all input parameters evaluated at level l-1.  We also improve on the hierarchical kriging \citep{Han2012} approach by making sure that not only information on the mean is incorporated between layers, but that information on the covariance is also shared. This is important as we should be also able to learn information about the covariance parameters as we increase up the hierarchical structure. 

The goal is to model the top level $f^{(L)}$ as a Gaussian process $Z^{(L)}$, given that there is valuable information stored in all of the lower levels.  We not only want to include information from level $L-1$ to model level $L$, but to make use of all information from all $L-1$ levels below. We also want to make sure that  information on both the form of the mean function, as well as the covariance function is included.  We hence define that level $l-1$ becomes a prior to level $l$ such that both the posterior predictive mean and covariance from level $l-1$ become prior forms for level $l$.

\section{Multi-level Gaussian Processes}

Let a single layer computer code or model be represented by the function $f : \mathcal{X} \mapsto \mathbb{R}$ with $\textbf{X} := (\textbf{x}_{1}, \ldots, \textbf{x}_{n})^{T} \in \mathcal{X} \subset \mathbb{R}^{p}$. The function $f(\cdot)$ maps a set of $n$ inputs in $p$ dimensions to the output given by $\textbf{y} = (f(\textbf{x}_{1}), \ldots, f(\textbf{x}_{n}))^{T}$.  We then assume that the code has been run at a set of input parameter settings $X$. We can model $f$ as a Gaussian process $Z$ with training data $\{\textbf{X}, \textbf{y}\}$ such that:
\begin{equation}
Z(\textbf{x}) \sim GP(m(\textbf{x}), k(\textbf{x}, \textbf{x}')),
\end{equation}
where $m(\cdot)$ and $k(\cdot, \cdot)$ are predefined mean and covariance functions of the GP.  Typically,  and without loss to generality, the mean function takes either a constant or linear form, $m(\textbf{x}) = h(\textbf{x})^{T}\beta$, such that $h(\textbf{x})$ is a vector of regression functions, whilst $\beta$ is a vector of regression parameters. Common forms of covariance function chosen are either the Matern or squared exponential:
\begin{equation}
k(\textbf{x}, \textbf{x}') = \sigma^{2} \text{exp} \left\{ -\sum_{j=1}^{p} \left( \frac{x_{j} - x_{j}'}{\delta_{j}} \right)^{2} \right\}.
\end{equation}
Based on these prior choices, a Gaussian process is hence defined by a set of hyperparameters $\theta = (\beta, \sigma^{2}, \delta)$, which must be optimised during the GP fitting.

Using standard multivariate Normal distribution results, we can predict the function $f$ at a new location $\textbf{x}$ with posterior mean and covariance given as: 
\begin{equation} \label{PMC}
\begin{split}
m^{*}(\textbf{x}) &= m(\textbf{x}) + k(\textbf{x},\textbf{X}) \textbf{K}^{-1} (\textbf{y} - m(\textbf{X})), \\
k^{*}(\textbf{x}, \textbf{x}') &= k(\textbf{x}, \textbf{x}') - k(\textbf{x},\textbf{X}) \textbf{K}^{-1} k(\textbf{X},\textbf{x}'),
\end{split}
\end{equation}
where $\textbf{K}$ is an $n \times n$ covariance matrix with entries $K_{ij} = k(\textbf{x}_{i}, \textbf{x}_{j})$.

\subsection{Multi-level computer models}

Now assume that the computer model $f$ can be run at different levels of fidelity. Let $l  = 1, \ldots, L$ represent each of the $L$ levels of the multi-level computer model, with the $l^{th}$ level denoted by $f^{(l)}$,  such that $l=1$ represents the most coarse or cheapest level, and $l=L$ represents the full complex or top level. For each level of the model, let $\textbf{X}^{(l)} := ( \textbf{x}_{1}^{(l)}, \ldots, \textbf{x}_{n_{l}}^{(l)})$ be a set of $n_{l}$ inputs and $\textbf{y}^{(l)} = (f^{(l)}(\textbf{x}_{1}^{(l)}), \ldots, f^{(l)}(\textbf{x}_{n_{l}}^{(l)}))$ be the vector of associated outputs.  

\subsection{Kennedy and O'Hagan multi-level Gaussian process}
The most common method for dealing with complex computer models with multiple levels of complexity was introduced by \cite{Kennedy2000}.  With the top level, $f^{(L)}(\textbf{x})$, being the function of interest, they employ an autoregressive model framework with the following Markov-like property:
\begin{equation}
\text{cov} \left[ f^{(l)}(\textbf{x}), f^{(l-1)}(\textbf{x}') | f^{(l-1)}(\textbf{x}) \right] = 0.
\end{equation}
This states that given we know the computer model at level $l-1$ at input parameter $x$, $f^{(l)}(\textbf{x})$, then we can't learn any more about $f^{(l)}(\textbf{x})$ from any other model run $f^{(l-1)}('textbf{x}')$ for $\textbf{x}' \neq \textbf{x}$.

The Markov property along with stationarity of each model level, the $l^{th}$ level of the model can be stated as:
\begin{equation} \label{Eq1}
f^{(l)}(\textbf{x}) = \rho^{(l-1)} \cdot f^{(l-1)}(\textbf{x}) + \underbrace{\left( f^{(l)}(\textbf{x}) - \rho^{(l-1)} \cdot f^{(l-1)}(\textbf{x}) \right)}_{:= \delta^{(l)}(\textbf{x})}.
\end{equation}
Here, $\delta^{(l)}(\textbf{x}) = f^{(l)}(\textbf{x}) - \rho^{(l-1)} \cdot f^{(l-1)}(\textbf{x})$, for $l=2, \ldots, L$, acts as a corrector to the $(l-1)^{\text{th}}$ level to model the difference between the two neighbouring levels, and $\rho^{(l-1)}$ acts as a correlation parameter controlling how much $f^{(l-1)}$ relates to $f^{(l)}$, such that if $\rho^{(l-1)} = 0$, then $f^{(l-1)}$ and $f^{(l)}$ are independent. 

Due to the independence between $f^{(l)}$ for $l=1,\ldots, L$,  both $f^{(1)}$ and each difference term $\delta^{(l)}$ for $l > 1$ can be modelled as a Gaussian process $Z^{(l)}$.   The associated training data is $D^{(l)} = \{ \textbf{X}^{(l)}, \textbf{y}^{(l)}\}$ with $\textbf{y}^{(l)} = \big(f^{(1)}(\textbf{x}_{1}^{(1)}), \ldots, f^{(1)}(\textbf{x}_{n_{1}}^{(1)}) \big)$ for $l=1$ and $\textbf{y}^{(l)} =  \big( \delta^{(l)}(\textbf{x}_{1}^{(l)}), \ldots, \delta^{(l)}(\textbf{x}_{n_{l}}^{(l)}) \big)$ for $l>1$. $Z^{(l)}(\textbf{x})$ for $l>1$ models $\delta^{(l)}$ independently to $Z^{(l-1)}$ with independent hyperparameters.

\section{Bayesian Hierarchical Gaussian Process} \label{GLcase}

We will now introduce the proposed Bayesian hierarchical Gaussian process model. The general case is first presented, with the case when there are two model levels presented in Section 3.1. Examples are given in Section 4.

First assume that the computer model $f$ can be run at $l=1,\ldots,L$ levels of fidelity such that $f^{(1)}$ represents the cheapest level and $f^{(L)}$ represents the most complex and expensive level.  Also assume that initial data $\{\textbf{X}^{(l)}, \textbf{y}^{(l)} \}$ is collected from each model level where $\textbf{X}^{(l)} = (\textbf{x}_{1}^{(l)}, \ldots, \textbf{x}_{n_{l}}^{(l)})$ and $\textbf{y}^{(l)} = (f^{(l)}(\textbf{x}_{1}^{(l)}), \ldots,  f^{(l)}(\textbf{x}_{n_{l}}^{(l)}))$ for $l \in \{1, \ldots, L\}$. 

In the model structure, level $l-1$ becomes a prior to level $l$ such that both the posterior predictive mean and covariance from level $l-1$ become prior forms for level $l$. By making this prior assumption, all levels $l$ of the model for $l=2, \ldots, L$ are conditional on the previous $l-1$ levels by assuming that each $y^{(l)}$ has a Gaussian process prior defined by:
\begin{equation}
\textbf{y}^{(l)} \sim \mathcal{N} \Big(m_{l-1} \big(\textbf{X}^{(l)} \big),  k_{l-1} \big(\textbf{X}^{(l)},\textbf{X}^{(l)} \big) \Big)
\end{equation}
where 
\begin{align} \label{levell}
m_{l-1} \big( \textbf{X}^{(l)} \big) &= m^{(l)}(\textbf{X}^{(l)}) + k^{(l)} \big(\textbf{X}^{(l)},\textbf{X}^{(l-1)} \big) k^{(l)}\big(\textbf{X}^{(l-1)},\textbf{X}^{(l-1)} \big)^{-1} \big(\textbf{y}^{(l-1)} - m^{(l)} \big(\textbf{X}^{(l-1)} \big) \big) \\
k_{l-1} \big( \textbf{X}^{(l)},\textbf{X}^{(l)} \big) &= k^{(l)}(\textbf{X}^{(l)}, \textbf{X}^{(l)}) - k^{(l)} \big(\textbf{X}^{(l)},\textbf{X}^{(l-1)} \big)  k^{(l)}\big(\textbf{X}^{(l-1)},\textbf{X}^{(l-1)} \big)^{-1}  k^{(l)} \big(\textbf{X}^{(l-1)},\textbf{X}^{(l)} \big).
\end{align}
Here, $m^{(l)}$ and $k^{(l)}$ are also predefined mean and covariance functions respectively, specific for level $l$. Both functions are dependent on the set of hyperparameters $\theta_{l-1}$ which are distinct from $\theta_{l-2}$. The form for posterior mean and variance for level $l$ at location $\textbf{x}$ are hence given as:
\begin{equation}
\begin{split}
m^{*}_{l}(\textbf{x}) &= m_{l-1}(\textbf{x}) + k_{l-1} \big(\textbf{x},\textbf{X}^{(l)} \big) k_{l-1} \big( \textbf{X}^{(l)}, \textbf{X}^{(l)} \big) ^{-1} \big(\textbf{y}^{(l)} - m_{l-1} \big(\textbf{X}^{(l)} \big) \big) \\
k^{*}_{l}(\textbf{x}, \textbf{x}') &=  k_{l-1}(\textbf{x}, \textbf{x}') - k_{l-1} \big(\textbf{x},\textbf{X}^{(l)} \big) k_{l-1} \big( \textbf{X}^{(l)}, \textbf{X}^{(l)} \big)^{-1} k_{l-1} \big(\textbf{X}^{(l)},\textbf{x}' \big).
\end{split}
\end{equation}

An appropriate simplification of Equation \eqref{levell} can be made when both $m_{l-1}$ and $k_{l-1}$ are taken to be the posterior distributions from level $l-1$.  This is equivalent to both $m^{(l)}$ and $k^{(l)}$ taking the forms of the prior mean and prior covariance of level $l-1$. We can therefore state the following:
\begin{align}
m_{l-1} \big( \textbf{X}^{(l)}  \big) &= m^{*}_{l-1} \big( \textbf{X}^{(l)}  \big) \\
&= m_{l-2}(\textbf{X}^{(l)} ) + k_{l-2} \big(\textbf{X}^{(l)} ,\textbf{X}^{(l-1)} \big) k_{l-2} \big( \textbf{X}^{(l-1)}, \textbf{X}^{(l-1)} \big) ^{-1} \big(\textbf{y}^{(l-1)} - m_{l-2} \big(\textbf{X}^{(l-1)} \big) \big) \\
k_{l-1} \big( \textbf{X}^{(l)} ,\textbf{X}^{(l)}  \big) &= k^{*}_{l-1}  \big( \textbf{X}^{(l)} ,\textbf{X}^{(l)} \big) \\
&= k_{l-2}(\textbf{X}^{(l)} , \textbf{X}^{(l)} ) - k_{l-2} \big(\textbf{X}^{(l)} ,\textbf{X}^{(l-1)} \big) k_{l-2} \big( \textbf{X}^{(l-1)}, \textbf{X}^{(l-1)} \big)^{-1} k_{l-2} \big(\textbf{X}^{(l-1)},\textbf{X}^{(l)} \big)
\end{align}

We note that since hyperparameters are optimised to estimate the model at the top level $L$ of the model, it is not possible to consider the Gaussian process as valid at intermediate levels of the model. Any predictions made on these levels would not give accurate results and there would be a high level of error.

\subsection{Model with 2 Levels} \label{2levels}

The Bayesian hierarchical multi-level Gaussian process is now presented for a model with $2$ levels of fidelity, where a GP $Z^{(2)}$ can be constructed to the training data from all both levels of the model. Let $\{\textbf{X}^{(l)}, \textbf{y}^{(l)} \}$ be the initial data with $\textbf{X}^{(l)} = (\textbf{x}_{1}^{(l)}, \ldots, \textbf{x}_{n_{l}}^{(l)})$ and $\textbf{y}^{(l)} = (f^{(l)}(\textbf{x}_{1}^{(l)}, \ldots, \textbf{x}_{n_{l}}^{(l)})$ for $l \in \{1, 2\}$. 

We first assume that $\textbf{y}^{(1)}$ has a multivariate normal distribution defined by:
\begin{equation}
\textbf{y}^{(1)} \sim \mathcal{N} \Big(m_{0} \big(\textbf{X}^{(1)} \big),  k_{0} \big(\textbf{X}^{(1)},\textbf{X}^{(1)} \big) \Big)
\end{equation}
such that $m_{0}$ and $k_{0}$ are predefined mean and covariance functions respectively.  

We wish to fit a GP $Z^{(2)}$ to the entire data across both levels of the model.  First assume that $\textbf{y}^{(1)}$ has a multivariate normal distribution defined by:
\begin{equation}
\textbf{y}^{(1)} \sim \mathcal{N} \Big(m_{0} \big(\textbf{X}^{(1)} \big),  k_{0} \big(\textbf{X}^{(1)},\textbf{X}^{(1)} \big) \Big)
\end{equation}
such that $m_{0}$ and $k_{0}$ are predefined mean and covariance functions respectively.  As mentioned in the previous section, these can take a variety of forms, and we point the reader to \cite{Rasmussen2006} for more information. 

Given unknown hyperparameters, $\theta^{(1)}$, to both $m_{0}$ and $k_{0}$,  the posterior mean and covariance at a generic location $\textbf{x}$ conditioned on the initial data from level $l=1$ can be given by:
\begin{equation}
\begin{split}
m^{*}_{1}(\textbf{x}) &= m_{0}(\textbf{x}) + k_{0} \big(\textbf{x},\textbf{X}^{(1)} \big) k_{0}\big(\textbf{X}^{(1)},\textbf{X}^{(1)} \big)^{-1} \big(\textbf{y}^{(1)} - m_{0} \big(\textbf{X}^{(1)} \big) \big) \\
k_{1}^{*}(\textbf{x}, \textbf{x}') &=  k_{0}(\textbf{x}, \textbf{x}') - k_{0} \big(\textbf{x},\textbf{X}^{(1)} \big)  k_{0}\big(\textbf{X}^{(1)},\textbf{X}^{(1)} \big)^{-1}  k_{0} \big(\textbf{X}^{(1)},\textbf{x}' \big)
\end{split}
\end{equation}

Given, the GP model defined for level 1, we now wish to find a GP for level 2 fully conditioned on level 1. We hence assume that $\textbf{y}^{(2)}$ also has a multivariate normal distribution which is conditional on the posterior distribution of $\textbf{y}^{(1)}$ as follows:
\begin{equation} \label{L2P1}
\textbf{y}^{(2)} \sim \mathcal{N} \Big( m_{1}\big(\textbf{X}^{(2)} \big),  k_{1} \big(\textbf{X}^{(2)},\textbf{X}^{(2)} \big) \Big)
\end{equation}
where we can define both the prior mean and covariance to take the form of the posterior mean and covariance from level 1 as follows:
\begin{align} \label{L2P3}
m_{1} \big( \textbf{X}^{(2)} \big) &= m^{(2)}(\textbf{X}^{(2)}) + k^{(2)} \big(\textbf{X}^{(2)},\textbf{X}^{(1)} \big) k^{(2)}\big(\textbf{X}^{(1)},\textbf{X}^{(1)} \big)^{-1} \big(\textbf{y}^{(1)} - m^{(2)} \big(\textbf{X}^{(1)} \big) \big) \\
k_{1} \big( \textbf{X}^{(2)},\textbf{X}^{(2)} \big) &= k^{(2)}(\textbf{X}^{(2)}, \textbf{X}^{(2)}) - k^{(2)} \big(\textbf{X}^{(2)},\textbf{X}^{(1)} \big)  k^{(2)}\big(\textbf{X}^{(1)},\textbf{X}^{(1)} \big)^{-1}  k^{(2)} \big(\textbf{X}^{(1)},\textbf{X}^{(2)} \big).
\end{align}
The posterior mean and variance for level 1 become the prior distribution for level 2 such that the posterior mean and variance for level 2 are now conditioned on both the data at level 1 and the data at level 2.  

As discussed in Section \ref{GLcase}, it is possible to make simplifications such that the exact form of the posterior for level $1$ becomes the prior for level $2$.  Equation 2 is updated to take this into account by the following:
\begin{align} \label{L2P3}
m_{1} \big( \textbf{X}^{(2)} \big) &= m^{*}_{1} \big( \textbf{X}^{(2)} \big) \\
& = m_{0}(\textbf{X}^{(2)}) + k_{0} \big(\textbf{X}^{(2)},\textbf{X}^{(1)} \big) k_{0}\big(\textbf{X}^{(1)},\textbf{X}^{(1)} \big)^{-1} \big(\textbf{y}^{(1)} - m_{0} \big(\textbf{X}^{(1)} \big) \big) \\
k_{1} \big( \textbf{X}^{(2)},\textbf{X}^{(2)} \big) &= k^{*}_{1}  \big( \textbf{X}^{(2)},\textbf{X}^{(2)} \big) \\
&= k_{0}(\textbf{X}^{(2)}, \textbf{X}^{(2)}) - k_{0} \big(\textbf{X}^{(2)},\textbf{X}^{(1)} \big)  k_{0}\big(\textbf{X}^{(1)},\textbf{X}^{(1)} \big)^{-1}  k_{0} \big(\textbf{X}^{(1)},\textbf{X}^{(2)} \big),
\end{align}
Following this, the posterior mean and variance at unknown location $\textbf{x}$ is hence given as:
\begin{equation} 
\begin{split}
m_{2}^{*}(\textbf{x}) &= m_{1}(\textbf{x}) + k_{1} \big(\textbf{x},\textbf{X}^{(2)} \big) k_{1} \big(\textbf{X}^{(2)},\textbf{X}^{(2)} \big)^{-1} \big(\textbf{y}^{(2)} - m_{1} \big(\textbf{X}^{(2)} \big) \big) \\
k_{2}^{*}(\textbf{x}, \textbf{x}') &=  k_{1}(\textbf{x}, \textbf{x}') - k_{1} \big(\textbf{x},\textbf{X}^{(2)} \big) k_{1}\big( \textbf{X}^{(2)}, \textbf{X}^{(2)} \big)^{-1} k_{1} \big(\textbf{X}^{(2)},\textbf{x}' \big).
\end{split}
\end{equation}

As with a typical single level Gaussian process, the Bayesian hierarchical GP is fully defined by a single mean function and covariance function.  The only prior choices to be made are the mean and covariance functions in level 1, defined as $m_{0}$ and $k_{0}$ respectively. This then results in the entire model is dependent on only one set of hyperparameters, $\theta_{0}$, which are contained within these two functions. 

The main differences between Equation \eqref{L2P3} and Equation 2 is the introduction of the new mean and covariance functions, $m^{(2)}$ and $k^{(2)}$ respectively. The forms of these functions do not have to be the same as either $m_{0}$ or $k_{0}$ and are dependent on the set of hyperparameters $\theta_{1}$, which are independent from $\theta_{0}$. Equation \eqref{L2P3} shows how the prior for level 2 can still be conditioned on the data from level 1, whist introducing a greater level of flexibility.  Not only have a new set of hyperparameters been introduced, but we can also state that both $m^{(2)}$ and $k^{(2)}$ can take a different form to the mean and covariance function in level 1. 

It can be particularly useful to have different values of both the mean parameters and the correlation length parameter when the top level has more intricate areas in the output that can not be successfully captured in the lop level parameters.

\section{Examples and Comparisons}

In this section, we present several examples with different numbers of input dimensions and levels.  1) 2d example with 2 levels, 2) 2d example with 2 levels where the correlation between levels is either high or low, 3) three special case 1d examples with two levels, 4) 2d example with either 3 or 4 levels, 5) 4d park function, 5) high dimensional example with 22 input variables. The performance of our method (BayHEm) is compared against two alternatives: the Kennedy and O'Hagan multi-level Gaussian process (K\&O) \citep{Kennedy2000} and hierarchical kriging (HK) \citep{Han2012}.

The accuracy of the mean predictions are determined by calculating the root mean squared error (RMSE) across the top level $L$. Given a test set $D_{\tau} = \{ \textbf{X}_{\tau}, f^{(L)}(\textbf{X}_{\tau}) \}_{\tau=1}^{\tau=N}$, the RMSE is given by:
\begin{equation}
\text{RMSE} = \frac{\sqrt{\Big(\sum_{\tau=1}^{N}\big(m^{*}_{L}(\textbf{x}_{\tau}) - f^{(L)}(\textbf{x}_{\tau})\big)^{2}\Big)}}{N},
\end{equation}
where $m^{*}_{L}(\textbf{x}_{\tau})$ is the expected mean output of the multi-level GP at input $\textbf{x}_{\tau}$.  The test set is generated as a size 10000 random latin hypercube for all 2d and 4d examples, and 1000 equally spaced points in the 1d examples. 

\subsection{Example I}

\begin{figure}[ht]
\centering
\includegraphics[scale=0.4]{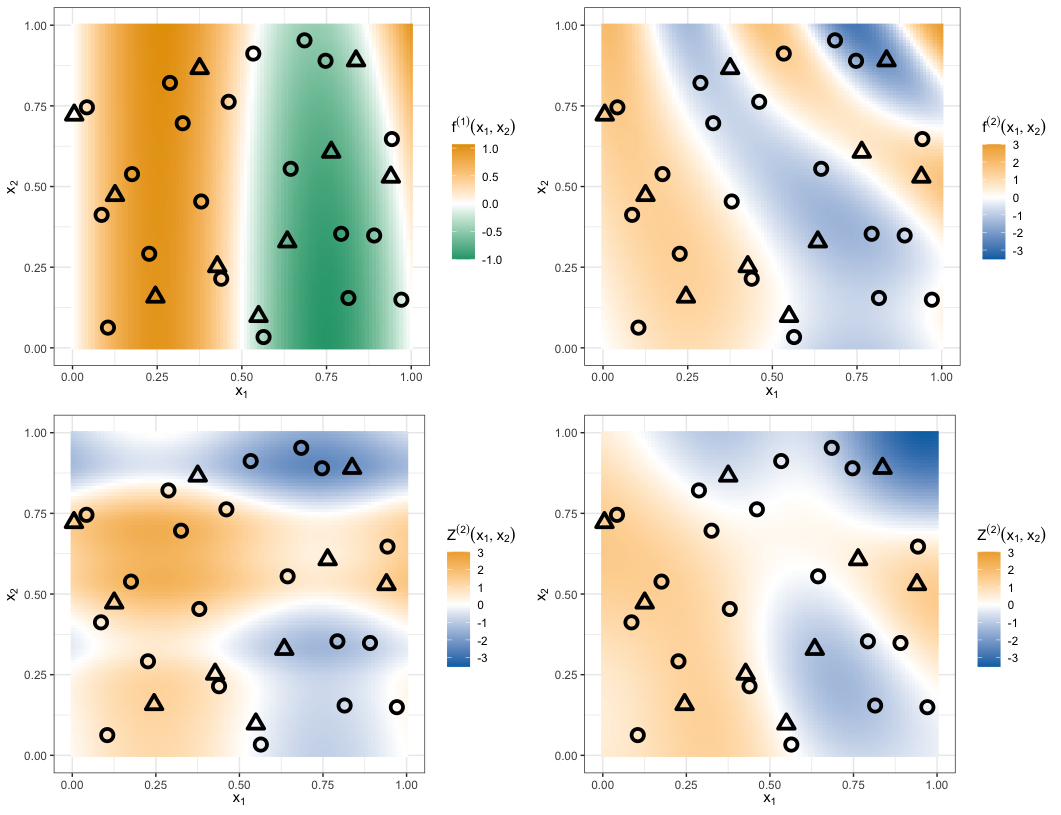}
\caption{-}
\label{Pic1}
\end{figure}

First consider a simple toy example with two levels.  There are two inputs to the model, $(x_{1}, x_{2}) \in \mathcal{X} = [0,1]^2$, with the two levels of the function given by:
\begin{align*}
f^{(1)}(x_{1}, x_{2}) &= (x_{1} x_{2})^{2} + \text{sin}(2 \pi x_{1}), \\
f^{(2)}(x_{1}, x_{2}) &= f^{(1)}(x_{1}, x_{2}) + 2 x_{2} \left( \text{cos}(4 \pi x_{1} x_{2}) + x_{1}^{2} - x_{1} x_{2} \right).
\end{align*}
Initial designs are generated using latin hypercubes with 20 points run at level 1 and 10 points run at level 2. The method from the previous section is applied and the results are given in Figure \ref{Pic1}. The top row show the true functions for both levels 1 and 2 respectively, along with initial points (circles for level 1 and triangles for level 2).  The bottom left plot shows the predicted mean output from applying the traditional Kennedy and O'Hagan approach, whilst the bottom right is the predicted mean from our new method.  We can notice that the higher values in our new method are better estimated, as well as the top right hand region in the plot. This is reflected in the RMSE scores being 1.1291 and 0.7471 respectively. Confirming that our method is performing better in this example.

The method is further evaluated by varying the ratio of level 1 to level 2 points. Keeping 20 points fixed at level 1, the number of level 2 points is adjusted to 20, 12, 10, and 5. Table 1 presents the RMSE scores for each combination across four emulator methods: (1) a single GP trained only on the top level, (2) the BayHEm GP, (3) the Kennedy and O’Hagan multi-level GP, and (4) hierarchical kriging. To mitigate the bias introduced by the experimental design, results are reported as the mean RMSE across 20 different Latin hypercube samples, with the range of RMSE values also provided.

BayHEm consistently achieves the lowest RMSE scores across all level 2 design sizes. Furthermore, it exhibits the smallest range of RMSE values, particularly in the upper bound, indicating stable and reliable predictions across different designs. Both the Kennedy and O’Hagan method and the single GP perform well when more points are available at level 2, but their accuracy declines as the number of level 2 points decreases. In contrast, hierarchical kriging maintains a relatively stable RMSE across all design sizes. While its average RMSE is comparable to BayHEm when the level 2 design contains only five points, its range of RMSE values is significantly larger, suggesting less consistent performance.

\begin{center}
\begin{tabular}{| c || c | c | c | c |} 
 \hline
 No. L2 points & Single GP & BayHEm  & K\&O & HK  \\ [0.5ex] 
 \hline\hline
 20 & 0.651 (0.55,0.76) & 0.574 (0.53,0.59) & 0.669 (0.57,0.85) & 0.745 (0.51,0.86) \\ 
 \hline
 12 & 0.988 (0.71,1.16) & 0.701 (0.60,0.84) & 0.887 (0.75,1.08) & 0.784 (0.68,0.85) \\
 \hline
 10 &  1.121 (0.89, 1.35) &  0.739 (0.63,0.85) & 1.008 (0.87,1.32) & 0.792 (0.72,0.85) \\
 \hline
 5 & 1.281 (0.89,1.38) & 0.808 (0.72,0.88) & 1.146 (0.88,1.34) & 0.851 (0.81,1.61) \\ [1ex] 
 \hline
\end{tabular} \label{Tabel1}
\end{center}

\begin{figure}[ht]
\centering
\includegraphics[scale=0.4]{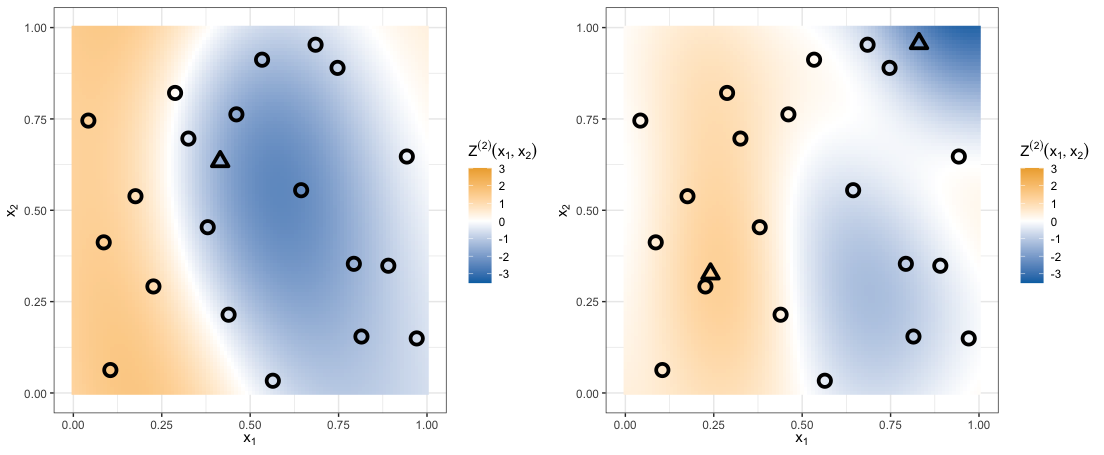}
\caption{-}
\label{Pic2}
\end{figure}

Figure 2 illustrates the performance of our proposed method when only a limited number of level 2 model runs are available. The left plot corresponds to a design with a single level 2 point, while the right plot includes two points. In both cases, the predicted mean is shown, demonstrating that the method effectively incorporates level 2 information even when data is extremely sparse.

\subsection{Example II: Correlation Between Levels}

The second example is designed to examine the effects of varying correlation between levels, considering both cases where they are highly correlated and where they are completely independent, with no shared information. While the Kennedy and O'Hagan method allows for the incorporation of a correlation parameter, we have chosen to fix it at 1 for experimental consistency. In the uncorrelated case, the parameter would be set to 0, reducing the multi-level GP to a standard single GP - a baseline method already included in our comparisons. To ensure a fair evaluation, we assume no prior knowledge of correlation across both examples and all methods. 

Both examples are in two dimensions with two levels. The levels of the correlated example are given by:
\begin{align*}
f^{(1)}(x_{1}, x_{2}) &= (x_{1} x_{2})^{2} + \text{sin}(2 \pi x_{1}) + \text{cos}(4 \pi x_{1} x_{2}), \\
f^{(2)}(x_{1}, x_{2}) &= (x_{1} x_{2})^{2} + \text{sin}(2 \pi x_{1}) +  2 x_{2} \left( \text{cos}(4 \pi x_{1} x_{2}) + x_{1}^{2} - x_{1} x_{2} \right).
\end{align*}
and the levels of the uncorrelated example are given by:
\begin{align*}
f^{(1)}(x_{1}, x_{2}) &= \frac{4 x_{1}^{3} - x_{1} x_{2}^{4}}{\text{exp}(x_{1} x_{2})^{2}} - 2, \\
f^{(2)}(x_{1}, x_{2}) &= (x_{1} x_{2})^{2} + \text{sin}(2 \pi x_{1}) +  2 x_{2} \left( \text{cos}(4 \pi x_{1} x_{2}) + x_{1}^{2} - x_{1} x_{2} \right).
\end{align*}

Similarly to Example I, the RMSE scores are presented in Table 2. When the two model levels are highly correlated, BayHEm and K\&O exhibit similar performance in predicting the mean response.  In the case where the levels are independent, the single GP, BayHEm, and hierarchical kriging yield similar mean RMSE scores and ranges. This demonstrates that both BayHEm and hierarchical kriging effectively recognise the absence of useful information in level 1 when modelling level 2. The highest RMSE is observed for K\&O, reflecting the challenge of estimating the difference between levels 1 and 2 with limited data. In this scenario, the function representing the difference between levels is more complex than the level 2 function itself.

\begin{center}
\begin{tabular}{| c || c | c | c | c |} 
 \hline
 Correlated & Single GP & BayHEm GP & K\&O GP & HK GP \\ [0.5ex] 
 \hline\hline
 Highly & 0.968 (0.77, 1.21) & 0.531 (0.41, 0.66) & 0.593 (0.47,0.68) & 0.737 (0.59, 0.92) \\ 
 \hline
 Uncorrelated & 0.902 (0.75, 1.11) & 0.861 (0.71, 1.08) & 1.181 (0.98, 1.25) & 0.895 (0.73,1.23) \\ [1ex] 
 \hline
\end{tabular}
\end{center}

\subsection{Example III: 1d special cases}

We now investigate a set of simple special cases in one dimension. Each of the three examples consists of a two-level model, where level 1 contains a large number of points (25), while level 2 is limited to only two points. The level 1 function is defined as:
\begin{equation}
f^{(1)}(x) = x \, \text{sin}(x) + x,
\end{equation}
whereas the three different level 2 functions are given by:
\begin{align*}
f^{(2)}_{1}(x) &= f^{(1)}(x) + 4, \\
f^{(2)}_{2}(x) &= f^{(1)}(x) + 2x, \\
f^{(2)}_{3}(x) &= f^{(1)}(x) + 4 \, \text{sin}\left(\frac{x}{2}\right).
\end{align*}

The first example represents a simple vertical shift between the two levels, while the second introduces the equivalent to a linear tilt along the x-axis at the origin. The third example transformation resembles a stretch along the y-axis. In this case, level 2 is greater than level 1 in the first half of the input space, whereas in the second half, level 1 exceeds level 2.

The results of fitting both the BayHEm GP and the K\&O multi-level GP are shown in Figure 3. The left-hand plots correspond to BayHEm, while the right-hand plots correspond to K\&O. The true level 1 function is depicted in blue, with training points represented by circles, while the true level 2 function is shown in orange, with its two training points marked by triangles. The predicted mean obtained from each GP method is represented by the green line, with uncertainty bounds indicated by grey dashed lines.

For the first example, the K\&O method demonstrates the best performance, yielding minimal error and uncertainty. This is expected, as the difference emulator in this case is simply estimating the constant function $f(x)=4$, making it well-suited for the example. While the BayHEm method also performs well, it does not match the accuracy of K\&O in this instance. However, in the second and third examples, the limitations of the K\&O method become apparent due to the limited amounts of level 2 data. In both cases, the predicted mean for level 2 closely follows the level 1 function, with only slight deviations towards the level 2 curve at the training points. In contrast, BayHEm demonstrates greater robustness under limited data conditions, providing a more accurate representation of the level 2 function.

\begin{figure}[ht]
\centering
\includegraphics[scale=0.3]{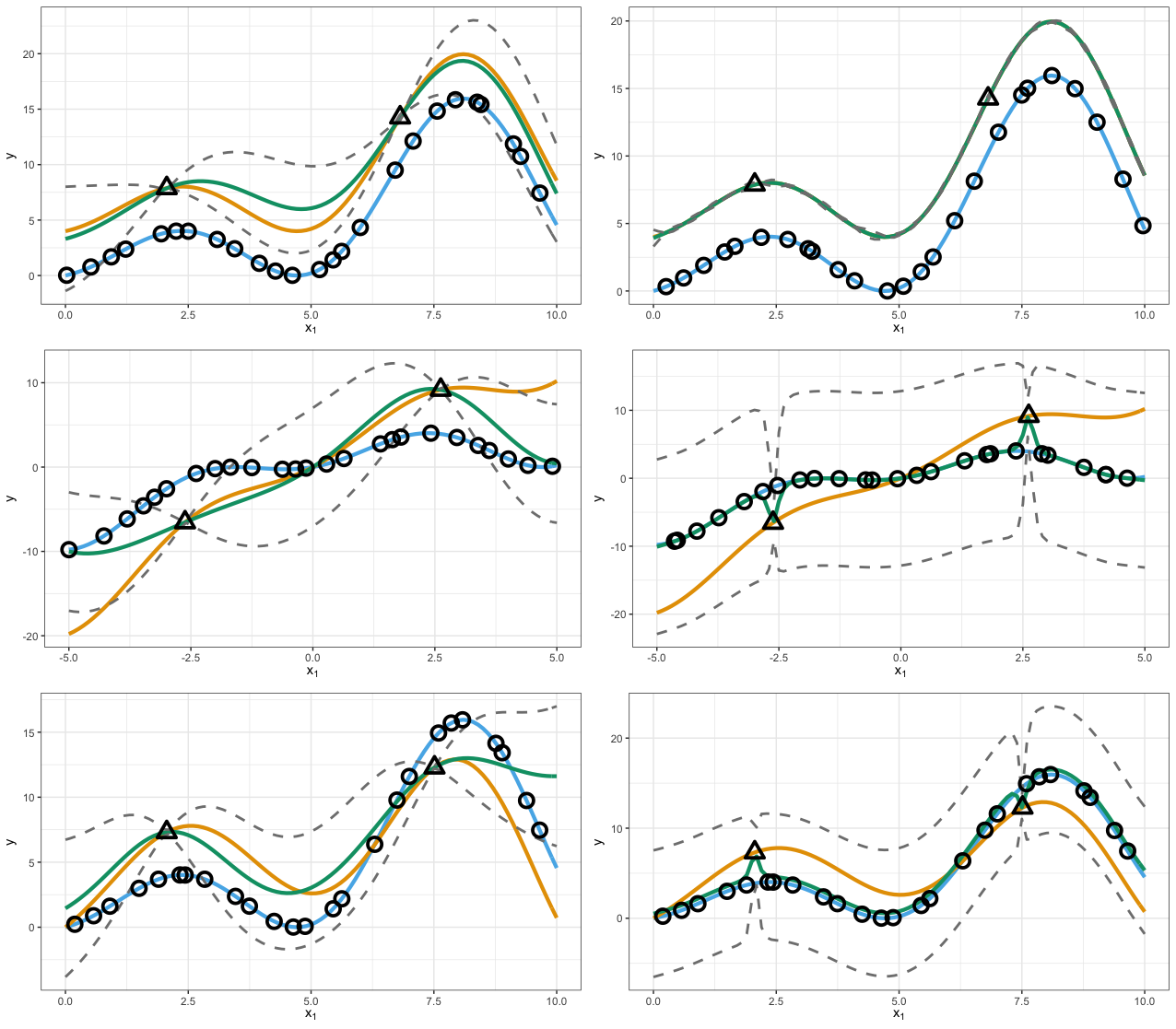}
\caption{-}
\label{Pic5b}
\end{figure}

\bibliographystyle{abbrvnat}
\bibliography{Bibfile1}


\end{document}